\overfullrule=0in
\magnification=\magstep1
\centerline{\bf SPHERICALLY-SYMMETRIC RANDOM WALKS IN NONINTEGER DIMENSION}
\bigskip
\bigskip
\centerline{Carl M. Bender}
\medskip
\centerline{Department of Physics}
\medskip
\centerline{Washington University}
\medskip
\centerline{St. Louis, Missouri 63130-4899~~USA}
\medskip
\centerline{and}
\medskip
\centerline{Stefan Boettcher}
\medskip
\centerline{Department of Physics}
\medskip
\centerline{Brookhaven National Laboratory}
\medskip
\centerline{Upton, New York 11973~~USA}
\medskip
\centerline{and}
\medskip
\centerline{Moshe Moshe}
\medskip
\centerline{Department of Physics}
\medskip
\centerline{Technion Israel Institute of Technology}
\medskip
\centerline{Haifa 32000~~ISRAEL}
\bigskip
\bigskip
\bigskip
\baselineskip=24pt
\centerline{\bf ABSTRACT}
\medskip
A previous paper proposed a new kind of random walk on a spherically-symmetric
lattice in arbitrary noninteger dimension $D$. Such a lattice avoids the
problems associated with a hypercubic lattice in noninteger dimension. This
paper examines the nature of spherically-symmetric random walks in detail. We
perform a large-time asymptotic analysis of these random walks and use the
results to determine the Hausdorff dimension of the process. We obtain exact
results in terms of Hurwitz functions (incomplete zeta functions) for the
probability of a walker going from one region of the spherical lattice to
another. Finally, we show that the probability that the paths of $K$ independent
random walkers will intersect vanishes in the continuum limit if $D>
{{2K}\over{K-1}}$.
\footnote{}{PACS numbers: 05.40.+j, 03.29+i, 02.50+5}
\vfill\eject

\noindent {\bf I. INTRODUCTION}
\bigskip

In a recent paper$^1$ we devised a $D$-dimensional spherical lattice on which
a random walk is well defined even when $D$ is noninteger. The advantage of such
a lattice is that for all real values of $D$ the probabilities have sensible and
acceptable values; that is, they lie between $0$ and $1$. A hypercubic lattice
gives acceptable probabilities only for integer $D$. Having constructed such a
model, our objective in the current paper is to examine the nature of spherical
random walks in much greater detail.

The simplest spherically-symmetric geometry is a set of regions bounded by 
concentric nested spheres. Such a geometry does not single out the integer
dimensions as being special in any way. Thus, a random walk can be defined in
any real dimension $D$. We consider an infinite set of concentric nested spheres
of radius $R_n$ ($n=1,~2,~3,~\ldots$). $Region~n$ is the volume between the
$n-1$st and the $n$th spheres (see Fig.~5 of Ref.~1). At each time step a random
walker in $region~n$ must move inward or outward to an adjacent region; the
probability of walking outward is denoted $P_{\rm out}(n)$ and the probability
of walking inward is $P_{\rm in}(n)$. In our model the probabilities
$P_{\rm out}(n)$ and $P_{\rm in}(n)$ are determined by the relative surface area
between adjacent regions. Thus,
$$\eqalign{P_{\rm out}(n)&={{S_D(R_n)}\over {S_D(R_n)+S_D(R_{n-1})}}~~,\cr
P_{\rm in}(n)&={{S_D(R_{n-1})}\over {S_D(R_n)+S_D(R_{n-1})}}~~,}\eqno{(1.1)}$$
where $S_D(R)$ is the surface area of a $D$-dimensional sphere of radius $R$:
$$S_D (R)={ {2\pi^{D/2}}\over{\Gamma (D/2)} } R^{D-1}~~.\eqno{(1.2)}$$
Since the surface area $S_D(R)$ is proportional to $R^{D-1}$, the inward-walk
and outward-walk probabilities are
$$\eqalign{P_{\rm out}(n)&={{R^{D-1}_n}\over {R^{D-1}_n+R^{D-1}_{n-1}}}~~,\cr
P_{\rm in}(n)&={{R^{D-1}_{n-1}}\over{R^{D-1}_n+R^{D-1}_{n-1}}}~~.}\eqno{(1.3)}$$
These formulas are valid for all $D$ and all $n\geq 2$. For $n=1$, $P_{\rm out}
(1)=1$ and $P_{\rm in}(1)=0$ because there is no region inward from $n=1$ and
therefore the walker {\sl must} go outward. Note that $P_{\rm out}$ and
$P_{\rm in}$ are nonnegative numbers less than or equal to $1$ with
$P_{\rm in} (n)+P_{\rm out} (n)=1$; thus, they are acceptable probabilities for
all real $D$.

A spherical random walk is described by the function $C_{n,t}$, which represents
the probability that a random walker arrives at $region~n$ at time~$t$. The
probability $C_{n,t}$ satisfies the recursion relation
$$\eqalign{C_{n,t}&=P_{\rm in}(n+1)C_{n+1,t-1}+P_{\rm out}(n-1)C_{n-1,t-1}
\quad (n\geq 2)~~,\cr C_{1,t} &= P_{\rm in}(2) C_{2,t-1}~~,}\eqno{(1.4)}$$
where $P_{\rm in}$ and $P_{\rm out}$ are given by (1.3). This equation states
that for a random walker to be in $region~n$ at time $t$, she must have come
from $region~n+1$ or $region~n-1$ at time $t-1$.

In Ref.~1 we considered the consequences of the particular initial condition
that at time $t=0$ the random walker is located in the central sphere
($region~1$). We represent this initial condition by
$$C_{n,0} = \delta_{n,1}~~.\eqno{(1.5)}$$

Two immediate consequences of (1.4) and (1.5) are that for all values of $t$, 
$$\sum_{n=1}^{\infty} C_{n,t}=1\eqno{(1.6)}$$
and that $C_{n,t}\geq 0$ for all $n\geq 1$ and $t\geq 0$. Hence, for all $n$
and $t$, $C_{n,t}$ lies between 0 and 1 and is thus an acceptable probability
for all real $D$.

In general, for arbitrary $D$ the solution $C_{n,t}$ to the partial difference
equation in (1.4) and (1.5) is rather complicated. However, in Ref.~1 we used
the theory of continued fractions to solve (1.4) and (1.5) for $C_{1,2t}$ for
any value of $D$ and for any sequence of radii $R_n$. We simply iterate (1.4)
and (1.5) to obtain
$$\eqalign{ C_{1,0}&= 1~~,\cr   C_{1,2}&= Q_1~~,\cr   C_{1,4}&=Q_1 (Q_1+Q_2 )~~,
\cr C_{1,6}&= Q_1 [ (Q_1+Q_2)^2 + Q_2Q_3 ]~~,} \eqno{(1.7)}$$
and so on, where
$$Q_n \equiv P_{\rm in} (n+1) P_{\rm out} (n)~~. \eqno{(1.8)}$$
The sequence of formulas in (1.7) is a known pattern:$^2$ $C_{1,2t}$ are the
coefficients in the Taylor series for the function whose continued-fraction
representation has the coefficients $Q_n$:
$$\sum_{t=0}^{\infty} C_{1,2t} x^t = 1/(1-Q_1 x/(1 - Q_2 x/(1 - Q_3 x/(
1 - Q_4 x/ \ldots ))))~~.\eqno{(1.9)}$$
Thus, at $x=1$ we have the formal result
$$\sum_{t=0}^{\infty}C_{1,2t}=1/(1-Q_1/(1-Q_2/(1-Q_3/(1-Q_4/\ldots))))~~.
\eqno{(1.10)}$$

If we truncate the continued fraction in (1.10) after $Q_n$ and substitute
$$\eqalign{ Q_1&={ {R_1^{D-1}}\over{R_1^{D-1}+R_2^{D-1}} }~~,\cr
Q_n&={ {R_n^{2D-2}}\over{(R_{n-1}^{D-1}+R_n^{D-1})(R_n^{D-1}+R_{n+1}^{D-1})}}
\quad (n\geq 2)~~,} \eqno{(1.11)}$$
an amazing algebraic simplification occurs for each value of $n$:
$$1/(1-Q_1/(1-Q_2/(1-Q_3/(1-Q_4/\ldots(1-Q_n)\ldots ))))=
\sum_{k=1}^n \left ( {{R_1}\over{R_N}} \right )^{D-1}~~.\eqno{(1.12)}$$
We may then take the limit $n\to\infty$ if the series in (1.12) converges. We
obtain
$$\sum_{t=0}^{\infty} C_{1,2t}=\sum_{k=1}^{\infty}\left ({{R_1}\over{R_k}}
\right ) ^{D-1} ~~.\eqno{(1.13)}$$
 For the special case $R_n=n$, corresponding to equally-spaced concentric
spheres$^3$, and $D>2$, (1.13) becomes
$$\sum_{t=0}^{\infty} C_{1,2t}=\zeta (D-1) ~~,\eqno{(1.14)}$$
where $\zeta$ is the Riemann zeta function. Note that the sum in (1.14) is
{\sl not} a probability because the probabilities $C_{1,2t}$ are not disjoint.
Nevertheless, the sum does have a physical interpretation. It is the {\sl mean
time} spent by the random walker in $region~1$.$^4$

Another interesting probability function is $P_{2t}$ ($t=1,~2,~3,~\ldots$), the
probability that the random walker returns to the central region ($region~1$)
at time $2t$ {\sl for the first time.} The generating-function formula$^4$
$$\sum_{t=1}^{\infty} P_{2t} x^t =1-{1\over {\sum_{t=0}^{\infty}
C_{1,2t} x^t }}\eqno{(1.15)}$$
states the relationship between $C_{1,2t}$ and $P_{2t}$. The generating-function
formula in (1.15) is universal; it holds for a random walk in any geometry.

If we sum over all disjoint (mutually exclusive) probabilities $P_{2t}$ we
obtain $\Pi_D$, the probability that a random walker will eventually return to
her starting point in $region~1$:
$$\Pi_D \equiv \sum_{t=1}^{\infty} P_{2t}~~.\eqno{(1.16)}$$
To compute $\Pi_D$ we set $x=1$ in (1.15) and use (1.10) and (1.13):
$$\eqalign{\Pi_D &=Q_1/(1-Q_2/(1-Q_3/(1-Q_4/(1-Q_5/\ldots))))\cr &=1- {1\over{
\sum_{N=1}^{\infty}\left ({{R_1}\over{R_N}}\right )^{D-1}}}~~.}\eqno{(1.17)}$$
Thus, for the special case in which the random walk occurs on a geometry of
concentric spheres whose radii are consecutive integers, $R_n=n$, we have
$$\eqalign{\Pi_D &= 1\quad (D\leq 2)~~,\cr
\Pi_D &= 1 - {1\over {\zeta (D-1)}} \quad (D>2)~~.}\eqno{(1.18)}$$

In this paper we extend the analysis of the partial difference equation (1.4).
In Sec.~II we perform a detailed asymptotic analysis of (1.4) for large $t$. We
use the results of this analysis in Sec.~III to compute the spatial moments of
$C_{n,t}$. The spatial moments represent the random walker's mean distance from
the origin at time $t$. From the spatial moments we can show that the Hausdorff
dimension of a spherical random walk is 2 for all $D>0$. In Sec.~IV we show how
to calculate the temporal moments. This calculation allows us to determine the
{\sl average time} it takes for a random walker to return to the central region.
In Sec.~V we generalize $\Pi_D$ to a quantity $\Pi^{m,n}_D$, the probability
that a random walker who begins in $region~n$ will eventually reach $region~m$
($n>m$). We show that $\Pi^{m,n}_D$ can be expressed in closed form in terms of
a Hurwitz function (an incomplete zeta function). Finally, in Sec.~VI we examine
the problem of intersecting random walks. We show that in the continuum limit
the probability that $K$ independent random walks will intersect vanishes when
the dimension $D$ exceeds $2K/(K-1)$. This reproduces the value of the critical
dimension for which a scalar $\phi^{2K}$ quantum field theory becomes free.

\vfill \eject
\noindent {\bf II. LARGE-TIME ASYMPTOTIC ANALYSIS OF SPHERICAL RANDOM WALKS}
\bigskip

The partial difference equation (1.4) augmented by the initial condition (1.5)
determines the function $C_{n,t}$ that describes a random walk beginning in the
central region of a spherically-symmetric lattice. In this section we perform a
large-time asymptotic analysis of this function $C_{n,t}$ for the flat-space
case $R_n=n$. Substituting $R_n=n$ into (1.4) gives the partial difference
equation
$$\eqalign{C_{n,t}&={{n^{D-1}}\over{(n+1)^{D-1}+n^{D-1}}}C_{n+1,t-1}+
{{(n-1)^{D-1}}\over{(n-1)^{D-1}+(n-2)^{D-1}} }C_{n-1,t-1} \quad (n\geq 3)~~,\cr
C_{2,t}&={{2^{D-1}}\over{3^{D-1}+2^{D-1}}}C_{3,t-1}+C_{1,t-1}~~,\cr
C_{1,t}&={1\over{1+2^{D-1}}}C_{2,t-1}~~.}\eqno{(2.1)}$$

 For arbitrary $D$ these equations do not have a closed-form solution. However,
we showed in Ref.~1 that for $D=1$ and $D=2$ there are simple, exact
expressions for $C_{n,t}$. For $D=1$,
$$\eqalign{C_{n,n+2j-1} &={{(n+2j-1)!}\over { (n+j-1)! j! 2^{n+2j-2}}}\quad
(n\geq 2)~~,\cr C_{1,2t} &= {{(2t)!}\over {t!t!2^{2t}}}~~.}\eqno{(2.2)}$$
 For $D=2$,
$$C_{n,n+2j-1}={{(2n-1)(n+2j-1)!}\over{j!(2n+2j-1)!!2^j}}~~,\eqno{(2.3)}$$
from which we have
$$C_{1,2t} = {1\over {1+2t}}~~.\eqno{(2.4)}$$

We begin by examining the large-time behavior of $C_{n,t}$ for the dimensions
$D=1$ and $D=2$ using the above exact solutions. We can expand the expressions
in (2.2) and (2.3) for large $t$ using the following asymptotic relation
satisfied by the Gamma function,
$${{\Gamma(t+\alpha)}\over{\Gamma(t)}}\sim t^{\alpha}\left [ 1+{{\alpha^2
-\alpha}\over {2t}}+{\rm O}(t^{-2})\right ]\quad (t\to\infty)~~,\eqno{(2.5)}$$
which may be derived from the Stirling expansion.$^5$ We obtain for $D=1$
$$\eqalign{C_{n,t}&\sim {{2^{3/2}}\over{\sqrt{\pi t}}}\left [ 1-{{2n^2-4n+3}
\over{4t}}+{\rm O}(t^{-2})\right ]\quad (t\to\infty,~n\geq 2)~~,\cr
C_{1,2t}&\sim {1\over{\sqrt{\pi t}}}\left [ 1-{1\over{8t}}+{\rm O}(t^{-2})
\right ]\quad (t\to\infty)}\eqno{(2.6)}$$
and for $D=2$
$$C_{n,t}\sim {{2n-1}\over t}\left [ 1-{{n^2-n+1}\over{2t}}+{\rm O}(t^{-2})
\right]\quad (t\to\infty)~~.\eqno{(2.7)}$$

Note that these asymptotic behaviors appear to have the form of Frobenius$^5$
series about $t=\infty$. That is, they have the form of Taylor series in inverse
powers of $t$ multiplied by a fractional power of $t$. This suggests that we
might seek an expression for $C_{n,t}$ of the form
$$C_{n,t}=t^{\alpha}{\cal C}(n)\sum_{k=0}^{\infty}c_k(n) t^{-k}~~,\eqno{(2.8)}$$
where we choose $c_0(n)\equiv 1$. Actually, detailed numerical calculations for
$C_{1,2t}$ out to $2t=200,000$ show that the formula in (2.8) is not quite
correct for arbitrary $D$; there appear to be logarithmic corrections which
become negligible when $n$ is large compared with $1$. In the following analysis
the parameter $n$ will be sufficiently large that we may ignore these
logarithmic terms.

 Frobenius series are particularly appropriate for the local analysis of
differential equations because the derivative of $t^{\alpha-k}$ with respect to
$t$ gives another term of the same form with the index $k$ shifted by one.
However, they are not suitable for difference equations because the {\sl
discrete difference} in the $t$ variable of the expression $t^{\alpha-k}$,
$$(t)^{\alpha-k} - (t-1)^{\alpha-k}~~,$$
is not simply expressible as a combination of terms of the form $t^{\alpha-k}$.
As is shown in Ref.~5, an appropriate form for a Frobenius series valid in the
vicinity of $t=\infty$ and useful for difference equations is
$$C_{n,t}={\cal C}(n)\sum_{k=0}^{\infty}c_k(n)
{{\Gamma(t+\alpha-k)}\over{\Gamma(t)}}~~,\eqno{(2.9)}$$
where $c_0(n)\equiv 1$. Note that we have replaced $t^{\alpha - k}$ in (2.7) by
$\Gamma(t+\alpha-k)/\Gamma(t)$. This {\sl ansatz} has two advantages. First, for
large $t$, we have
$$\Gamma(t+\alpha-k)/\Gamma(t) \sim t^{\alpha - k}~~,\eqno{(2.10)}$$
by virtue of Eq.~(2.5). Second, the discrete difference of $\Gamma(t+\alpha-k)/
\Gamma(t)$ is simply expressible in terms of the same structure with $k$ shifted
by one:
$${{\Gamma(t+\alpha-k)}\over{\Gamma(t)}}
- {{\Gamma(t-1+\alpha-k)}\over{\Gamma(t-1)}}
= (\alpha - k) {{\Gamma(t+\alpha-k-1)}\over{\Gamma(t)}}~~.\eqno{(2.11)}$$
Indeed, (2.11) is the discrete analog of the derivative relation
$${{d}\over{dt}} t^{\alpha - k} = (\alpha - k) t^{\alpha - k -1}~~.$$

Our program now is to substitute (2.9) into the partial difference equation
(2.1). To do so we make use of the identity
$$C_{n,t-1}={\cal C}(n)\sum_{k=0}^{\infty}c_k(n){{\Gamma(t+\alpha-k)}\over
{\Gamma(t)}}+{\cal C}(n)\sum_{k=1}^{\infty}(k-\alpha-1)c_{k-1}(n){{\Gamma(t
+\alpha-k)}\over{\Gamma(t)}}~~.\eqno{(2.12)}$$
Then we compare coefficients of $\Gamma(t+\alpha-k)/\Gamma(t)$. This comparison
gives a recursion relation to be solved for ${\cal C}(n)$ and $c_k(n)$ ($n\geq
1$). There are three cases to consider, $D>1$, $0\leq D\leq 1$, and $D<0$. 

\leftline{\bf Case 1: $D>1$}

This is the simplest case because the three equations in (2.1) reduce to a
single equation valid for all $n\geq 1$; there are no special equations for
$n=1$ and $n=2$. The $k=0$ equation determines the coefficient ${\cal C}(n)$:
$${\cal C}(n)={{n^{D-1}}\over{(n+1)^{D-1}+n^{D-1}}}{\cal C}(n+1)+
{{(n-1)^{D-1}}\over{(n-1)^{D-1}+(n-2)^{D-1}}}{\cal C}(n-1)~~.\eqno{(2.13)}$$
The solution to this equation is
$${\cal C}(n) = A [n^{D-1}+(n-1)^{D-1}]~~,\eqno{(2.14)}$$
where $A$ is a constant that cannot be determined by (2.13) because the
equation is homogeneous.

The $k\geq 1$ equations yield a recursion formula that gives the coefficient
$c_k(n)$ in terms of $c_{k-1}(n)$:
$$[n^{D-1}+(n-1)^{D-1}]c_k(n)-n^{D-1}c_k(n+1)-(n-1)^{D-1}c_k(n-1)$$
$$=(k-\alpha-1)[n^{D-1} c_{k-1}(n+1)+(n-1)^{D-1} c_{k-1}(n-1)~~.\eqno{(2.15)}$$

 From the first of these equations, $k=1$, we obtain a formula for $c_1(n)$:
$$c_1(n) = c_1(1) + \alpha\sum_{m=1}^{n-1} {1\over {m^{D-1}}}
\sum_{j=1}^m [j^{D-1} + (j-1)^{D-1}]~~.\eqno{(2.16)}$$
Note that the solution for $c_1(n)$ contains the arbitrary summation constant
$c_1(1)$ and the (as yet undetermined) index $\alpha$.

The remaining equations express $c_k(n)$ as a sum over $c_{k-1}$:
$$c_k(n) = c_k(1) + (\alpha+1-k)\sum_{m=1}^{n-1} {1\over {m^{D-1}}}
\sum_{j=1}^m [j^{D-1}c_{k-1}(j+1)+(j-1)^{D-1}c_{k-1}(j-1)]~~.\eqno{(2.17)}$$
 For each value of $k$ we have a new arbitrary constant; we expect an infinite
number of arbitrary constants because we are solving a {\sl partial} difference
equation. In principle, these constants can be determined by a boundary
condition. However, no such boundary condition is available to us because we are
performing a {\sl local} asymptotic analysis at $t=\infty$. Ultimately, we will
be interested in the large-$n$ behavior of $c_k(n)$ and thus these arbitrary
constants will prove to be irrelevant.

We will now determine the value of the index $\alpha$. We do so by examining the
behavior of $c_k(n)$ for large $n$. We already have
$$c_0(n) \equiv 1~~.\eqno{(2.18a)}$$
Next, we evaluate $c_1(n)$ in (2.16) for large $n$. The upper regions of the
sums dominate in this asymptotic approximation and the result is {\sl
independent of the constant} $c_1(1)$:
$$c_1(n) \sim {{\alpha}\over D} n^2 \quad (n>>1)~~.\eqno{(2.18b)}$$
Using the result in (2.18b) we find the large-$n$ behavior of $c_2(n)$ from
(2.16); again, the upper regions of the sums dominate and the result is
independent of the constant $c_2(1)$:
$$c_2(n) \sim {{\alpha (\alpha -1)}\over {D(D+2)}} {{n^4}\over 2} \quad
(n>>1)~~.\eqno{(2.18c)}$$
Iterating this process we find that
$$c_3(n) \sim {{\alpha (\alpha -1) (\alpha -2)}\over {D(D+2) (D+4)}} {{n^6}\over
6} \quad (n>>1)~~.\eqno{(2.18d)}$$
In general we obtain the result
$$c_k(n)\sim{{\alpha (\alpha -1)(\alpha -2)\cdots (\alpha -k+1)}\over{D(D+2)
(D+4)\cdots (D+2k-2)}}{{n^{2k}}\over {k!}} \quad (n>>1)~~.\eqno{(2.18e)}$$
Equations (2.18) can be summarized as
$$c_k(n)\sim{{(-1)^k \Gamma(D/2) \Gamma(k-\alpha) n^{2k}}\over{2^k k!
\Gamma(k+D/2) \Gamma(-\alpha )}} \quad (n>>1)~~.\eqno{(2.19)}$$
If we set 
$$\alpha = -D/2\eqno{(2.20)}$$
then (2.19) greatly simplifies:
$$c_k(n)\sim{{(-1)^k n^{2k}}\over{2^k k!}}\quad (n>>1)~~.\eqno{(2.21)}$$
Note that in the above asymptotic analysis the variable $n$ cannot exceed
${\rm O}(\sqrt{t})$. This is because we are performing an asymptotic expansion
valid for large $t$. Subsequent terms in the expansion (2.9) must be of
decreasing size for our analysis to be correct.

We will now argue that (2.20) is in fact the correct value of the index
$\alpha$. We substitute (2.10), (2.14), and (2.21) into the discrete Frobenius
series (2.9) and obtain for large $t$,
$$C_{n,t}\sim 2A n^{D-1} \sum_{k=0}^{\infty} t^{-k-D/2} {{(-n^2)^k}\over
{2^k k!}}\quad [t\to\infty,~1<<n\leq {\rm O}(\sqrt{t})]~~,$$
which sums to
$$C_{n,t} \sim 2A n^{D-1} t^{-D/2} e^{-n^2/(2t)} \quad (t\to\infty,~1<<n)~~,
\eqno{(2.22)}$$
where the upper asymptotic limit on $n$ is no longer necessary. Observe that as
a function of the variable $n$, $C_{n,t}$ rises until it reaches a maximum at
$$n=\sqrt{(D-1)t}\eqno{(2.23)}$$
and then decays to zero exponentially. To illustrate this maximum we have
computed $C_{n,t}$ numerically for $D=3$ from (1.4-5) and graphed the results in
 Fig.~1 as a three-dimensional plot. Observe that the crest follows a parabola.
Note also that the maximum is very broad.

We now impose the normalization condition in (1.6), which states that at each
time $t$ the probability that the random walker is somewhere on the lattice is
unity. Substituting (2.22) into (1.6) gives
$$1=\sum_{n=1}^{\infty} 2A (2n)^{D-1} t^{-D/2} e^{-2n^2/t}~~,\eqno{(2.24)}$$
where we have assumed that $t$ is {\sl odd} and thus we must sum over {\sl even}
values of $n$ only. We could just as well take $t$ to be even and to sum over
odd $n$ only.$^6$ Since the dominant contribution to this sum comes from large
values of $n$ near $\sqrt{(D-1)t}$ we need not be concerned about $C_{n,t}$ for
small $n$ and we may immediately replace the sum in (2.24) by an integral over
$n$:
$$1=\int_{n=0}^{\infty} dn\; 2A (2n)^{D-1} t^{-D/2} e^{-2n^2/t}~~.$$
We evaluate this integral and obtain a value for the constant $A$:
$$A = {2\over {2^{D/2} \Gamma(D/2)}}~~.\eqno{(2.25)}$$
It is crucial that this constant be independent of the value of $t$; we find
that in evaluating the integral over $n$ the variable $t$ {\sl scales out} and
all $t$ dependence does indeed disappear. The requirement that the result be
independent of $t$ is precisely the condition that determines the value of the
index $\alpha$ in (2.20). Using the value of $A$ in (2.25) and taking just the
first term in the series (2.9) we obtain the leading asymptotic behavior of
$C_{n,t}$ for large $t$:
$$C_{n,t}\sim{2\over{2^{D/2}\Gamma(D/2)}}[n^{D-1}+(n-1)^{D-1}]t^{-D/2}
\quad (t\to\infty)~~.\eqno{(2.26)}$$

The result in (2.26) agrees exactly with the leading asymptotic behavior in
(2.7) for $C_{n,t}$ at $D=2$. It also reproduces exactly the leading asymptotic
behavior in (2.6) for $C_{n,t}$ at $D=1$. (To obtain the correct result at $n=1$
we must be careful to set $n=1$ first and then to allow $D$ to approach 1 from
above.) We have verified the behavior in (2.26) numerically by computing
$C_{n,t}$ for various values of $D$ out to $t=500$ and we find that our
asymptotic approximation agrees with the exact result to many decimal places.

\leftline{\bf Case 2: $0<D\leq 1$}

The analysis of Case 2 is more difficult than that of Case 1. If we formally
assume a Frobenius series (2.9) and substitute this series into the partial
difference equation (2.1), then as in Case 1, we can determine the large-$n$
behavior of the series coefficients. We substitute the Frobenius series into the
normalization condition (1.6) and, as with Case 1, we convert the sum in (1.6)
to an integral that is independent of $t$ provided that the index satisfies
$\alpha=-D/2$, the same result as that in (2.20).

When $D\leq 1$ we encounter a problem not present in the analysis of Case 1.
Recall from equation (2.23) that when $D>1$ the function $C_{n,t}$ increases
with $n$ until it reaches a crest at $n=\sqrt{(D-1)t}$ and then it decays. This
crest follows a parabolic curve in the $(n,t)$-plane. The crest structure
justifies replacing a sum in (2.24) by an integral, which we then evaluate to
determine the constant $A$ exactly. In Case 2, this crest still exists but it
lies along a straight line near $n=1$. Thus, the large-$n$ asymptotic behavior
of the Frobenius series coefficients is not sufficient to determine the value of
the constant $A$. To illustrate the change in the crest structure we have
calculated $C_{n,t}$ numerically for $D=1/2$ and plotted the results in Fig.~2.
Observe that unlike the situation shown in Fig.~1, the crest is parallel to the
$t$ axis.

Although we cannot {\sl derive} the coefficient of the leading asymptotic
behavior of $C_{n,t}$ for large $t$ we will make the assumption that we can
analytically continue the result in (2.25) from values of $D$ greater than $1$
(Case 1) to values of $D$ less than $1$ (Case 2). Of course, the analytic
continuation of an asymptotic approximation is not always the same as the
asymptotic approximation to the analytic continuation of a function.
Nevertheless, it is worth examining numerically the possibility that the leading
asymptotic behavior of $C_{n,t}$ for large $t$ is given by (2.26) (supplemented
by the special condition that the behavior of $C_{1,t}$ is obtained by first
setting $n=1$ and {\sl then} continuing $D$ below $1$). To test this hypothesis
we have computed $C_{n,t}$ for all values of $t$ up to $3000$ for $D=1/2$. In
 Fig.~3 we compare the exact value of $C_{1,t}$ with the leading asymptotic
approximation to this function taken from (2.26):
$$C_{1,t} \sim {{2^{3/4}}\over {\Gamma(1/4)}}t^{-1/4}~~.\eqno{(2.27)}$$
Clearly, (2.27) is an extremely good approximation for large $t$.

\leftline{\bf Special Case: $D=0$}

 For this special case there is no simple formula for $C_{n,t}$;
indeed, the numbers $C_{1,2t}$ are quite complicated looking:
$$\eqalign{ C_{1,0} &= 1~~,\cr C_{1,2} &= {{2} \over {3}}~~,\cr
C_{1,4} &= {{26} \over {45}}~~,\cr C_{1,6} &= {{502} \over {945}}~~,\cr
C_{1,8} &= {{7102}\over {14175}}~~,\cr C_{1,10} &= {{44834} \over {93555}}~~.}$$
Nevertheless, we have found an exact formula for $P_{2t}$ that is extremely
simple:$^1$
$$P_{2t} = {2\over {(2t+1)(2t-1)}}~~.\eqno{(2.28)}$$

Although it is probably valid to continue (2.26) to values of $D$ less than $1$,
this continuation clearly cannot be valid at $D=0$ because $\Gamma(D/2)$ becomes
divergent. Note that as $D$ approaches $0$ the algebraic behavior in the
variable $t$ disappears; also, the coefficient in (2.26) vanishes because
$\Gamma(0)=\infty$. This suggests to us that at $D=0$, $C_{n,t}$ continues to
vanish in the limit as $t\to\infty$ but that it vanishes less rapidly than
algebraically. In fact, $C_{n,t}$ exhibits a logarithmic decay as $t\to\infty$.

It is easy to see why there is a logarithmic decay in $C_{n,t}$ for large $t$.
If we assume that
$$C_{n,t} \sim {\cal F}(t) {\cal C}(n)\quad (t\to\infty)$$
and we substitute this behavior into (2.1) we obtain a formula for ${\cal C}(n)$
that is the exact analog of (2.14):
$${\cal C}(n)=\left ({1\over n}+{1\over{n-1}}\right ){\cal C}(1)\quad (n>1)~~.
\eqno{(2.29)}$$
To find ${\cal C}(1)$ we use the normalization condition in (1.6). Note that the
sum over $n$ in this case would be infinite except for the fact that $C_{n,t}$
must vanish for $n>t+1$. Thus, summing over $n$ with an upper limit of $t$ gives
the crude estimate that ${\cal F}(t)\sim (\ln t)^{-1}$ as $t\to\infty$. This
analysis is not delicate enough to give the overall multiplicative constant in
the asymptotic behavior.

A more precise result can be obtained by solving (1.15) for $\sum_{t=0}^{\infty}
x^t C_{1,2t}$ and substituting the exact formula for $P_{2t}$ in (2.28). We
obtain
$$\sum_{t=0}^{\infty} x^t C_{1,2t}={{2\sqrt{x}}\over{(1-x)
\ln[(1-\sqrt{x})/(1+\sqrt{x})]}}~~.\eqno{(2.30)}$$
Next we multiply (2.30) by $x^{-k-1}$ and integrate around a contour enclosing
the origin in the $x$-plane. We can evaluate this contour integral using the
method of steepest descents.$^5$ A saddle point occurs at $x=1$. We find
that the large-t asymptotic behavior is
$$C_{1,2t} \sim {2\over\ln 2t}\quad (t\to\infty)~~.\eqno{(2.31)}$$
We have verified this result numerically for values of $t$ up to $6000$.

\leftline{\bf Case 3: $D<0$}

When $D<0$ we find an entirely new behavior for $C_{n,t}$ for large $t$;
namely, that $C_{n,t}$ approaches a nonvanishing steady-state function of $n$:
$$\lim_{t\to\infty} C_{n,t} = C_{n,\infty}~~.\eqno{(2.32)}$$
To determine the function $C_{n,\infty}$ we merely set $t=\infty$ in (2.1) and
solve the steady-state equation:
$$\eqalign{C_{n,\infty}&={{n^{D-1}}\over{(n+1)^{D-1}+n^{D-1}}}C_{n+1,\infty}+
{{(n-1)^{D-1}}\over{(n-1)^{D-1}+(n-2)^{D-1}}}C_{n-1,\infty}\quad (n\geq 3)~~,\cr
C_{2,\infty}&={{2^{D-1}}\over{3^{D-1}+2^{D-1}}}C_{3,\infty}+C_{1,\infty}~~,\cr
C_{1,\infty}&={1\over{1+2^{D-1}}}C_{2,\infty}~~.}\eqno{(2.33)}$$
The solution to (2.33) is uniquely determined up to a single multiplicative
constant $C_{1,\infty}$:
$$C_{n,\infty}=[n^{D-1}+(n-1)^{D-1}]C_{1,\infty}\quad (n\geq 2)~~.
\eqno{(2.34)}$$

To find the value of $C_{1,\infty}$ we normalize the solution in (2.34) using 
(1.6) in the limit $t\to\infty$. Recall that the solution $C_{n,t}$ is nonzero
in a checkerboard pattern.$^6$ Thus, we can take $t$ to be odd, in which case
(1.6) becomes
$$\sum_{n=1}^{\infty} C_{2n,t} = 1~~,$$
or we may take $t$ to be even, in which case
$$\sum_{n=0}^{\infty} C_{2n+1,t} = 1~~.$$
In either of these two cases, after we substitute (2.34) and let $t\to\infty$,
the normalization condition becomes
$$(1^{D-1}+ 2^{D-1}+ 3^{D-1}+ 4^{D-1}+\cdots) C_{1,\infty}=1~~,$$
whence
$$C_{1,\infty} = {1\over {\zeta(1-D)}}\quad (D<0)~~.\eqno{(2.35)}$$
Thus, for large $t$ we have$^6$
$$\eqalign{C_{n,t} &\sim{1\over{\zeta(1-D)}}[n^{D-1}+(n-1)^{D-1}]\quad
(t\to\infty,~n\geq 2)~~,\cr
C_{1,t} &\sim {1\over {\zeta(1-D)}}\quad (t\to\infty)~~.}\eqno{(2.36)}$$

To illustrate the result in (2.36) we present a three-dimensional plot of the
function $C_{n,t}$ in Fig.~4 for $D=-1$. Observe that as $t\to\infty$, $C_{n,t}$
approaches a time-independent distribution.

An alternative way to obtain the value of $C_{1,\infty}$ makes use of a crucial
symmetry property of the partial difference equation (1.4) referred to in Ref.~1
as the {\sl reciprocity} property of (1.4). It was noted in Ref.~1 that if the
value of $D$ is reflected about the point $D=1$ then $P_{\rm out}(n)$ and
$P_{\rm in}(n)$ in (1.3) interchange roles when $n\geq 2$. From this observation
it was shown from the properties of continued fractions$^7$ that
$$P_{2t}\Bigm |_{D}=(C_{1,2t-2}-C_{1,2t})\Bigm |_{2-D}
\quad(t\geq 1)~~,\eqno{(2.37)}$$
where $P_{2t}$ is defined in (1.15). An immediate consequence of this property
can be obtained by summing (2.37) from $t=1$ to $\infty$:
$$\eqalign{ \Pi_D &=\sum_{t=1}^{\infty} P_{2t}\Bigm |_{D}\cr
&=(C_{1,0} - C_{1,\infty})\Bigm |_{2-D}\cr
&= 1- C_{1,\infty} \Bigm |_{2-D} ~~.}\eqno{(2.38)}$$

Hence, $\Pi_D=1$ if and only if $\lim_{t\to\infty}C_{1,2t}=0$ in $(2-D)$
dimensions. Thus, the fact that in Cases 1 and 2 we have $C_{n,t}\to 0$ as
$t\to\infty$ when $D>0$ implies that $\Pi_D=1$ for $D<2$. However, we have
computed $\Pi_D$ explicitly [see (1.17)] for $D>2$. Thus, using (2.38) we
reproduce the value of now $C_{1,\infty}$ given in (2.35).

To obtain higher-order corrections to the leading asymptotic behavior in (2.36)
we assume that the difference between $C_{n,t}$ and its asymptotic behavior in
(2.36) has the form of the discrete Frobenius series on the right side of
(2.9). (This procedure will be successful because, as we have verified
numerically, the difference between (2.36) and the exact result has a crest
as in Case 1.) We proceed exactly as in Case 1. Again we find that ${\cal C}(n)$
is determined up to an unknown constant $A$:
$$\eqalign{{\cal C}(n) &= A[n^{D-1} + (n-1)^{D-1}]\quad (n\geq 2)~~,\cr
{\cal C}(1) &= A~~.}\eqno{(2.39)}$$
Note that this equation is slightly different in form from that in (2.14)
because $D\leq 0$.

Next we find a recursive formula for the coefficients $c_k(n)$:
$$\eqalign{c_k(n)=c_k(1)  &-(\alpha+1-k)\sum_{m=1}^{n-1}c_{k-1}(m)\cr
&+(\alpha+1-k)\sum_{m=1}^{n-1}m^{1-D}\sum_{j=1}^m j^{D-1}[c_{k-1}(j)+c_{k-1}
(j+1)]~~.}\eqno{(2.40)}$$
This formula allows us to determine $c_k(n)$ for large $n$. From our
assumption that
$$c_0(n) \equiv 1~~,\eqno{(2.41a)}$$
we have
$$c_1(n)\sim{{2\alpha}\over {2-D}}\zeta(1-D) n^{2-D}\quad (n>>1)~~,
\eqno{(2.41b)}$$
$$c_2(n)\sim {{2\alpha (\alpha -1)}\over {(2-D)(4-D)}}\zeta(1-D){{n^{4-D}}
\over {1!}} \quad (n>>1)~~,\eqno{(2.41c)}$$
$$c_3(n)\sim{{2\alpha (\alpha -1)(\alpha -2)}\over{(2-D)(4-D)(6-D)}}
\zeta(1-D) {{n^{6-D}}\over {2!}} \quad (n>>1)~~,\eqno{(2.41d)}$$
and so on.

We will now show that the choice of index
$$\alpha = D/2~~,\eqno{(2.42)}$$
which we have found from numerical computations, is consistent with the
normalization condition (1.6). [When $D<0$ the normalization condition does not
uniquely specify the value of $\alpha$ but we can at least show that the choice
of $\alpha$ in (2.42) does not contradict the normalization condition.] If we
substitute this value of $\alpha$ into (2.41), we find that
$$c_k(n) \sim { {\zeta(1-D) D n^{2k-D}}\over {(2k-D)(-2)^{k-1}(k-1)!} }\quad
(k\geq 1,~n>>1)~~.\eqno{(2.43)}$$
[Note that in this equation the variable $n$ should not exceed ${\rm O}
(\sqrt{t})$. See (2.21).] Thus, for large $t$ the first correction to $C_{n,
\infty}$, apart from the multiplicative function ${\cal C}(n)$, is
$$t^{D/2}+D \zeta(1-D) \sum_{k=1}^{\infty}{{t^{-k+D/2} n^{2k-D}}
\over{(2k-D)(-2)^{k-1}(k-1)!}}~~,\eqno{(2.44)}$$
where we have substituted the asymptotic result in (2.43) into the Frobenius
series in (2.9) and taken $t$ large. The sum in (2.44) can be represented as an
integral
$$t^{D/2} + D\zeta(1-D) t^{-1+D/2} n^{2-D} \int_0^1 dx\; x^{1-D}
e^{x^2n^2/(2t)}~~.\eqno{(2.45)}$$

We now multiply the result in (2.45) by ${\cal C}(n)$ in (2.39) and sum over all
even $n$ (or all odd $n$). It is justifiable to replace the sum over $n$ by an
integral for each fixed $x$ in the integral in (2.45) because there is a
crest for large $n$. The result of doing the $n$ integration is
$$At^{D/2}\zeta(1-D)\left( 1+D\int_0^1 dx\; x^{-1-D}\right )=0~~.\eqno{(2.46)}$$
This shows that there are {\sl no} small time-dependent corrections to the
normalization condition in (1.6).

\leftline{\bf Summary of the large-time Asymptotic Behavior of $C_{n,t}$}

In this section we have investigated the asymptotic behavior of $C_{n,t}$ as
$t\to\infty$. We can summarize this behavior as follows:
$$C_{n,t} \sim {\cal C}(n) t^{\alpha} \quad (t\to\infty)~~.$$
When $D>0$, we find that $\alpha=-D/2$. Furthermore, for $D>1$, we have
determined the coefficient of $t^{\alpha}$:
$${\cal C}(n) = {2\over{2^{D/2}\Gamma(D/2)}}[n^{D-1}+(n-1)^{D-1}]~~.$$
When $D<0$, we find that $\alpha=0$ and thus $C_{n,t}$ approaches a
time-independent function of $n$ as $t\to\infty$; this function is given in
(2.36). Moreover, we have argued that the difference between $C_{n,t}$ and its
limiting value at $t=\infty$ is a function that falls off with increasing $t$
like $t^{D/2}$. We have verified this last result using detailed computer
calculations.

The special case $D=0$ has a logarithmic behavior for large $t$:
$$C_{n,t}\sim \left ({1\over n}+{1\over{n-1}}\right ){2\over{\ln t}}
\quad (t\to\infty)~~.$$

 From the large-$t$ asymptotic behavior of $C_{1,2t}$ we can determine the
large-$t$ asymptotic behavior of $P_{2t}$ defined in (1.15) using the
reciprocity relation in (2.37). This asymptotic behavior takes the general form
$$P_{2t}\sim {\cal P}t^{\beta}\quad (t\to\infty)~~.\eqno{(2.47)}$$
 For all values of $D\neq 2$ we derive from (2.37) that
$$\beta = -1 -{{|D-2|}\over 2}~~.\eqno{(2.48)}$$

This result is definitely {\sl not} true when $D=2$ because $\Pi_2$ exists; it
is clear that there must be logarithmic corrections in the asymptotic behavior
of $P_{2t}$ for this special value of $D$. Indeed from (2.31) we see that
$$P_{2t}\sim {2\over {t [\ln (2t)]^2}}\quad (t\to\infty)~~.$$

When $D>2$ the coefficient ${\cal P}$ is not known. However, when $D<2$ we have
$${\cal P}={{(2-D) 2^{D-2}}\over{\Gamma(1-D/2)}}~~.\eqno{(2.49)}$$ 

The coefficient in (2.49) can be verified for the case $D=0$. From (2.28) we see
that for large $t$, $P_{2t}\sim t^{-2}/2$, which agrees with (2.47-49).
Equations (2.47-49) are also valid at $D=1$. It was shown in Ref.~1 that at
$D=1$
$$P_{2t} = {{(2t)!}\over {t!t!(2t-1) 2^{2t}}}~~.\eqno{(2.50)}$$
 For large $t$ (2.50) becomes
$$P_{2t} \sim {1\over {2\sqrt{\pi}t^{3/2}}}~~,\eqno{(2.51)}$$
which agrees with (2.47-49).

\vfill\eject
\noindent {\bf III. SPATIAL MOMENTS OF RANDOM WALKS}
\bigskip

In the previous section we calculated the large-$t$ asymptotic behavior of the
solution to the partial difference equation (2.1) that describes a random walk
on a set of concentric spherical shells of radii $R_n=n$. This random walk is
subject to the initial condition (1.5) that states that the random walker begins
in the central region. In this section we use the results of this asymptotic
analysis to determine the mean distance of the random walker from the central
region as a function of time $t$ for large $t$.

The $k$th spatial moment of a random walk is defined as a weighted average over
the probabilities $C_{n,t}$:
$$\langle R^k\rangle_t \equiv \sum_{n=1}^{\infty} n^k C_{n,t}~~.\eqno{(3.1)}$$

To evaluate the sum in (3.1) we substitute the asymptotic behavior of $C_{n,t}$,
assume that $t$ is odd and sum over even $n$ (or equivalently we may assume
that $t$ is even and sum over odd $n$). The leading large-$t$ asymptotic
behavior of $C_{n,t}$ is completely known for $D>1$ [see (2.26)]. We thus have
$$\langle R^k\rangle_t \sim {{4t^{-D/2}} \over {\Gamma(D/2) 2^{D/2}}}
\sum_{n=1}^{\infty} (2n)^{k+D-1} e^{-2n^2/t} \quad (t\to\infty)~~.\eqno{(3.2)}$$

We evaluate the sum over $n$ in (3.2) by converting it to an integral:
$$\langle R^k\rangle_t \sim {{4 t^{-D/2}} \over {\Gamma(D/2) 2^{D/2}}}
\int_0^{\infty} dn\; (2n)^{k+D-1} e^{-2n^2/t}\quad (t\to\infty)~~.\eqno{(3.3)}$$
Replacing the sum by an integral is justified when $D>1$ because the sum is
dominated by large values of $n$ lying near the parabola $n^2 = (D-1)t$. Next,
we evaluate the integral:
$$\langle R^k\rangle_t \sim {{\Gamma({{k+D}\over
2})}\over{\Gamma({D\over 2})}} (2t)^{k/2}\quad (t\to\infty)~~.\eqno{(3.4)}$$
Note that the power of $t$, which is a critical exponent, for large $t$ is
independent of $D$. Although the formula in (3.4) is derived assuming that
$D>1$, we believe that (3.4) remains valid for all $D>0$. We have verified
this assertion by means of detailed numerical calculations.

\leftline{\bf Hausdorff Dimension of a Random Walk}

The Hausdorff dimension $D_{\rm H}$ of a random walk is defined as the
reciprocal of the critical exponent of $\sqrt{\langle R^2\rangle_t}$, the
root-mean-square distance from the origin.$^8$ Thus, to determine the Hausdorff
dimension, we substitute the asymptotic approximation in (3.4) with $k=2$:
$$\sqrt{\langle R^2\rangle_t}\sim\sqrt{Dt}\quad (t\to\infty)~~.\eqno{(3.5)}$$
The critical exponent is $1/2$ so we conclude that for all dimensions $D>0$
we have the universal result
$$D_{\rm H}=2~~.\eqno{(3.6)}$$
This result is the same as for random walks on Euclidean hypercubic lattices.

\leftline{\bf Special Case $D=1$}

It is a bit surprising to assert that the dimension of a random walk is $2$ even
if the dimension of the space is less than $2$. To demonstrate that this is
indeed true, in this subsection we evaluate $D_H$ directly for the special case
$D=1$. When $D=1$, we have the exact result that for odd $t$
$$\langle R^2\rangle_t=1+t+{{4\Gamma({t\over 2}+1)}\over{\sqrt{\pi}\Gamma({{t+1}
\over 2})}}~~.\eqno{(3.7)}$$
Thus, $\langle R^2\rangle_t$ grows like $t$ for large values of $t$ and we
have $D_H=2$ at $D=1$.

To derive the result in (3.7) we use an integral representation of the inverse
Beta function$^9$ to give an integral representation for $c_{n,t}$ in (2.2):
$$c_{n,t} ={4\over\pi}\int_0^{\pi/2} dx \;(\cos{x})^t \cos{[(n-1)x]}$$
with the appropriate factor $1/2$ for $n=1$. It is important to note that
this integral does not exhibit $(n-t)$-parity nor does it vanish when
$n>t+1$. To compute $\langle R^2\rangle_t$ one must perform a finite summation
of a geometric series. The resulting trigonometric functions can then be
done explicitly using formulas like$^{10}$
$$1={2\over\pi}\int_0^{\pi/2} dx\; (\cos{x})^t {\sin{[(t+1)x]}\over\sin{x}}~~.$$

Note that when $D<0$ there is no $t$-dependence in the root-mean-square as
$t\to\infty$. The result for the root-mean-square is divergent for $D$ between
0 and -2; thus $D_H = 0$ in this region. When $D<-2$, $D_H = \infty$.

\vfill\eject
\noindent {\bf IV. TEMPORAL MOMENTS OF RANDOM WALKS}
\bigskip

The quantity $\Pi_D$, as defined in (1.16) represents the probability that a
random walker who begins in the central region will eventually return to the
central region. In Ref.~1 it was shown that when $R_n=n$, $\Pi_D$ can be
computed in closed form as a function of $D$ [see (1.18)]. The probability
$\Pi_D$ may be regarded as the {\sl zeroth temporal moment} of the discrete
probability distribution $P_{2t}$.

Another interesting quantity that characterizes the structure of random walks is
$T$, the {\sl mean time elapsed before the first return to the central region.}
The mean elapsed time $T$ is a weighted average of the quantities $P_{2t}$:
$$T\equiv {\displaystyle{{\sum_{t=0}^{\infty}2t P_{2t}}}\over\displaystyle{{
\sum_{t=0}^{\infty}P_{2t}}}}~~.
\eqno{(4.1)}$$
The quantity $T$ is the {\sl first temporal moment} of the probability function
$P_{2t}$ divided by $\Pi_D$.

Although the sum in (1.16) for $\Pi_D$ converges for all values of $D$, it is
clear from (2.47) and (2.48) that $T$, as defined in (4.1), converges only if
$D<0$ or $D>4$; $T$ is infinite when $D$ lies in the range $0\leq D\leq 4$. Note
that when $D\leq 2$, a random walker who begins in the central region will
eventually return to the central region with a probability of unity but that the
average time for this return is infinite! The reason for this apparently
paradoxical behavior is that $T$ is an {\sl average} over all random walks that
return to the central region; many of these random walks are infinitely long.
These long walks dominate the average and cause it to diverge. As $D$ increases
the shorter walks begin to dominate the first moment of $P_{2t}$; in higher
dimension the longer walks almost never find their way back to the central
region. Thus, it is not surprising that eventually $T$ becomes a finite
quantity. The result that $T$ is infinite when $0\leq D\leq 4$ is true of
Euclidean lattices (see Ref.~4).

When $D<0$ it is possible to calculate $T$ in closed form using the reciprocity
theorem. We multiply (2.37) by $2t$ and sum over $t$:
$$\eqalign{\sum_{t=1}^{\infty}2t P_{2t}\Bigm |_{D}
&=\sum_{t=1}^{\infty} 2t(C_{1,2t-2}-C_{1,2t})\Bigm |_{2-D}\cr
&=\sum_{t=1}^{\infty} [(2t-2)C_{1,2t-2}-2tC_{1,2t}+2C_{1,2t-2}]\Bigm |_{2-D}\cr
&=\lim_{t\to\infty}(-2tC_{1,2t}) \Bigm |_{2-D}+2\zeta(1-D)~~,}\eqno{(4.2)}$$
where we have used (1.14). Since $2-D>2$, then by the asymptotic behavior in
(2.26) the limit vanishes and we have the exact result
$$T=2\zeta(1-D)\quad (D<0)~~.\eqno{(4.3)}$$
[Note that we have used the result$^1$ that $\Pi_D=1$ for $D\leq 2$ to obtain
(4.3).]

To calculate $T$ when $D>4$ we use (1.15) to relate the first moment of
$P_{2t}$ to the first moment of $C_{1,2t}$:
$$\eqalign{ T &= {d\over{dx}}{
\displaystyle{{\sum_{t=0}^{\infty}x^{2t}P_{2t}}}\over{\Pi_D}}\Bigm |_{x=1}\cr
&={d\over{dx}}{{-1}\over\displaystyle{{\sum_{t=0}^{\infty}x^{2t}C_{1,2t}{\Pi_D}}
}} \Bigm |_{x=1}\cr &={\displaystyle{{\sum_{t=0}^{\infty} 2t C_{1,2t}}}\over
\displaystyle{{\left (\sum_{t=0}^{\infty} C_{1,2t}\right )^2 {\Pi_D}}}}\cr &={{
\displaystyle{\sum_{t=0}^{\infty} 2t C_{1,2t}}}\over{[\zeta(D-1) - 1] \zeta(D-1)
}}~~,}\eqno{(4.4)}$$
where we have used (1.14) and (1.18).

To evaluate the sum in the numerator of (4.4) we differentiate the
continued-fraction representation in (1.9). In general, of course, the
derivative of a continued-fraction representation is rather complicated.
However, for the specific continued fraction in (1.9) we find that if we first
truncate the continued fraction after the term $Q_n x$ and then differentiate
once with respect to $x$, we obtain a sequence of converging structures that we
easily recognize:
$$\sum_{t=0}^{\infty} t C_{1,2t} = \sum_{n=2}^{\infty}{{n-1}\over{n^{D-1}}}
+ \sum_{n=1}^{\infty} n^{D-1} \zeta(D-1,n)^2~~,\eqno{(4.5)}$$
where
$$\zeta(\alpha,n) \equiv \sum_{k=n+1}^{\infty}{1\over{k^{\alpha}}}\eqno{(4.6)}$$
is a Hurwitz function (incomplete zeta function). It is easy to show that the
expression converges for $D>4$ and diverges for $D\leq 4$ (it is logarithmically
divergent at $D=4$).

\vfill\eject
\noindent {\bf V. RANDOM WALKS FROM $region~n$ TO $region~m$}
\bigskip

Until now we have always considered random walks that begin in the central
region ($region~1$). However, to prepare for our consideration of intersecting
random walks in the next section we address in this section the problem of
random walks that originate in $region~n$. We will calculate here the
probability that a random walker from $region~n$ will eventually reach
$region~m$, where $m<n$; let $\Pi^{m,n}$ represent this probability.

We begin our discussion by considering the special case $m=1$ and calculate
$\Pi^{1,n}$, the probability that a random walk that begins in $region~n$
eventually reaches the central region. Note that there are two values of $n$
for which we already know the answer:
$$\Pi^{1,1}=1\eqno{(5.1a)}$$
and
$$\Pi^{1,2} =\Pi_D~~,\eqno{(5.1b)}$$
where $\Pi_D$ is given in (1.18). Equation (5.1a) holds because a random walk
that originates in $region~1$ has {\sl already} reached $region~1$. To
understand (5.1b) recall that $\Pi_D$ represents the probability that a random
walk that begins in $region~1$ will eventually {\sl return} to $region~1$. On
the first step of such a walk the random walker goes from $region~1$ to
$region~2$ with a probability of $1$ because $P_{\rm out}(1)=1$; now, the
probability that the walker reaches $region~1$ is $\Pi^{1,2}$.

Assume that the random walk begins in $region~n$. After one step, the walker
moves outward with a probability of $P_{\rm out}(n)$ to $region~n+1$ or inward
with a probability of $P_{\rm in}(n)$ to $region~n-1$. Hence the probability
that the walker eventually reaches $region~1$ is a sum of two terms:
$$\Pi^{1,n}=P_{\rm in}(n)\Pi^{1,n-1}+P_{\rm out}(n)\Pi^{1,n+1}~~.\eqno{(5.2)}$$
This equation is an ordinary linear second-order homogeneous difference equation
whose solution is uniquely determined by the two initial conditions in (5.1).
The form of the exact solution is particularly elementary:
$$\Pi^{1,n} = {\displaystyle{{\sum_{k=n}^{\infty}{ {R_1^{D-1}}\over{R_k^{D-1}}}}
}\over \displaystyle{{\sum_{k=1}^{\infty} {{R_1^{D-1}}\over{R_k^{D-1}}}}}}~~.
\eqno{(5.3)}$$

Now we consider the general case of a random walk from $region~n$ to $region~m$.
Again, we identify two initial conditions. Clearly,
$$\Pi^{m,m}=1\eqno{(5.4a)}$$
because a random walk starting in $region~m$ has already reached $region~m$.
Next, we calculate $\Pi^{m,m+1}$ following the same procedure used to calculate
$\Pi_D$ and we obtain the probability in the form of a continued fraction like
that in (1.17):
$$\Pi^{m,m+1} = P_{\rm in}(m+1) [1/(1-Q_{m+1}/(1-Q_{m+2}/(1-Q_{m+3}/(1-Q_{m+4}/
\ldots))))]~~.$$
We then convert this continued fraction to a series like that in (1.17):
$$\Pi^{m,m+1}=1-{1\over\displaystyle{{\sum_{k=m}^{\infty}{{R_m^{D-1}}
\over{R_k^{D-1}}}}}}~~.\eqno{(5.4b)}$$

Next we construct a difference equation satisfied by $\Pi^{m,n}$. We argue that 
in the first step of a random walk that starts in $region~n$ the walker either
moves outward with a probability of $P_{\rm out}(n)$ to $region~n+1$ or inward
with a probability of $P_{\rm in}(n)$ to $region~n-1$. Hence the probability
of reaching $region~m$ eventually is the sum of two probabilities:
$$\Pi^{m,n}=P_{\rm in}(n)\Pi^{m,n-1}+P_{\rm out}(n)\Pi^{m,n+1}~~.\eqno{(5.5)}$$
The exact unique solution to (5.5) satisfying the two initial conditions in
(5.4) is a trivial generalization of that in (5.3):
$$\Pi^{m,n} = {\displaystyle{{\sum_{k=n}^{\infty} {{R_1^{D-1}}\over{R_k^{D-1}}}}}\over\displaystyle{{\sum_{k=m}^{\infty} {{R_1^{D-1}}\over{R_k^{D-1}}}}}}~~,
\eqno{(5.6)}$$
assuming that $m\leq n$.

Note that if we take $R_n=n$ then when $D\leq 2$ the upper end of the sums in
(5.6) dominate and we have $\Pi^{m,n}=1$. When $D>2$, $\Pi^{m,n}$ is the ratio
of two incomplete zeta functions:
$$\Pi^{m,n} = {{\zeta(D-1,n-1)}\over{\zeta(D-1,m-1)}}~~,\eqno{(5.7)}$$
where $\zeta(\alpha,n)$ is defined in (4.6).

\vfill\eject
\noindent {\bf VI. INTERSECTING RANDOM WALKS}
\bigskip

In rigorous mathematical studies of self-interacting scalar quantum field theory
one defines a regulated version of the quantum field theory on a Euclidean
space-time lattice. One then obtains the continuum quantum field theory in the
subtle and difficult limit in which the lattice spacing tends to zero. In the
lattice version of the quantum field theory one can show that there is a deep
connection between random walks and self-interacting scalar quantum field theory
(see Ref.~4). Indeed, on the basis of studies of random walks, it has been shown
that a $\phi^{2K}$ quantum field theory becomes free when the dimension of
space-time exceeds ${2K}\over{K-1}$. Proofs that a quantum field theory is
noninteracting rely on studies of intersecting random walks.$^{11}$
Specifically, it has been shown that a $\phi^{2K}$ quantum field theory is free
whenever there is vanishing probability that $K$ independent random walks will
intersect.

Rigorous treatments of quantum field theory on a Euclidean lattice have the
shortcoming that the lattice can only be defined in integer dimension. (Thus,
while it is known that a $\phi^4$ theory is free in five or more dimensions and
interacting in three or less dimensions, it is not known if the theory is
free in four dimensions.) However, we have shown that on a lattice consisting of
nested hyperspheres one can define a random walk in any noninteger dimension.
Thus, if we construct a quantum field theory on such a lattice (as unrealistic
as such a field theory may be) we may be able to study the transition of this
field theory from interacting to free as a function of the space-time dimension.
The discussion of such a rotationally-symmetric quantum field theory is the
subject of a future paper. However, in this paper we address the crucial
question of the probability of intersection of independent random walks.

The random walks described in this paper are one dimensional; random walks occur
in the radial coordinate only. Thus, the intersection of two
rotationally-symmetric random-walk paths has a different meaning from the
intersection of two paths on a hypercubic lattice. It is intuitively clear,
however, that we can study the analog of the intersection of two paths provided
that appropriate weighting factors are introduced in order to take into account
the difference between the two kinds of random walks, as we will now argue.

On a hypercubic lattice one can define$^4$ the {\sl conditional probability}
$P({\bf x}_1,t,{\bf x}_0,t_0)$ for a random walker to be on site {\bf x} at time
$t$ starting from the initial position ${\bf x}_0$ at time $t_0$. Similarly, on
a rotationally-symmetric lattice we can define $C^{n,p}_t$ as the probability of
a walker starting at $t=0$ in $region~p$ to be in $region~n$ at time $t$. Note
that $C^{n,1}_t$ is the same as $C_{n,t}$ as defined in Sec.~1.

Roughly speaking, one can think of the probability $C_{n,t}$ as a sum over the
conditional probability $P({\bf x}_1,t,{\bf x}_0,t_0)$:
$$\eqalign{{\sum_{{{\bf x}_1\in region~n}\atop{{\bf x}_0\in region~1}}
P({\bf x}_1,t,{\bf x}_0,0)}}~~.\eqno{(6.1)}$$
Similarly, one can think of $C^{n,p}_t$ as the sum
$$\eqalign{{\sum_{{{\bf x}_1\in region~n} \atop{
{\bf x}_0\in region~p}}P({\bf x}_1,t,{\bf x}_0,0)}}~~.\eqno{(6.2)}$$

The probabilities $C^{n,p}_t$ are not mutually exclusive. Thus, the sum
$\sum_{t=p-n}^\infty C^{n,p}_t$, where we take $n\leq p$, is not itself a
probability; rather, it represents the mean time spent in the $n$th region.
This sum is finite only for $D>2$ [see, for example, Eq.~(1.14)]. One can
define${^4}$ $P_1({\bf x}_1,t,{\bf x}_0,t_0)$ as the probability for a walker on
a hypercubic lattice to be at site ${\bf x}$ at time $t$ {\sl for the first
time} starting from the initial position ${\bf x}_0$ at time $t_0$. Similarly,
$P^{n,p}_t$ represents the probability of a random walker starting in
$region~p$ to be in $region~n$ ($n\leq p$) at time $t$ {\sl for the first time}.
Roughly speaking, $P^{n,p}_t$ can be thought of as the sum
$$\eqalign{{\sum_{{{\bf x}_1\in region~n} \atop{{\bf x}_0\in region~p}}P_1({
\bf x}_1,t,{\bf x}_0,0)}}~~.\eqno{(6.3)}$$

The mutually exclusive probabilities $P_{2t}$ defined in (1.15) are given by 
$P_{2t}\equiv P^{1,2}_{2t-1}$. In contrast to $\sum_{t=p-n}^\infty C^{n,p}_t$,
the sum on $P^{n,p}_t$ {\sl is} a probability and in fact $\Pi^{n,p}$
($n\leq p$), as defined in Sec.~V, is given by
$$\eqalign{\sum_{t=p-n}^\infty P^{n,p}_t}=\Pi^{n,p}~~.\eqno{(6.4)}$$
The generalization of the generating-function formula (1.15) is found to be:
$$\eqalign{\sum_{t=p-n}^\infty P^{n,p}_t x^t}={{\sum_{ t=p-n}^\infty C^{n,p}_t
x^t}\over{\sum_{ t=0}^\infty C^{n,n}_t x^t}}~~.\eqno{(6.5)}$$
To obtain (1.15) from (6.5) note that $C^{1,1}_{t+1}=C^{1,2}_t$.

On a hypercubic lattice, the probability ${\cal P}({\bf x_1},{\bf x_2},\Lambda)$
that two random walk paths starting at ${\bf x}_1$ and ${\bf x}_2$
(${\bf x}_1\neq{\bf x}_2$) will intersect in a region $\Lambda$, satisfies$^4$
$$\eqalign{{\cal P}({\bf x}_1,{\bf x}_2,\Lambda)\leq\sum_{{\bf z}\in\Lambda}
\Pi({\bf z}-{\bf x}_1)\Pi({\bf z}-{\bf x}_2)}~~,\eqno{(6.6)}$$
where $\Pi({\bf x})={\sum_{t=0}^\infty P_1({\bf x},t)}$.

One is interested in the continuum limit of this expression. In order to
calculate the right side of (6.6) we use our results for the
rotationally-symmetric random walk probabilities $\Pi^{m,n}$ in (5.6) and we
note the difference between $\Pi({\bf z-x})$ and $\Pi^{n,p}$, which includes the
summation over the sites in the regions. The appropriate probability that must
be used in (6.6) is $\Pi^{n,p}$ multiplied by
$${\sum_t C^{n,n}_t\over\sum_t C^{1,1}_t}~~.$$
The right side (6.6) is thus given by:
$$\eqalign{\sum_{n=1}^N\Pi^{n,p}~\Pi^{n,q}~\biggl({\sum_t C^{n,n}_t\over
\sum_t C^{1,1}_t}\biggr)^2}~~.\eqno{(6.7)}$$

To obtain the continuum limit we now rescale all dimensions by
$$\eqalign{p\rightarrow\alpha p,\quad q\rightarrow\alpha q,\quad N\rightarrow
\alpha N }~~.\eqno{(6.8)}$$
 From (5.6) one notes that $\Pi^{n,p}~\sim~(p/n)^{2-D}$ and the correction
factor in (6.7) eliminates the $n$ dependence for large $n$ in the probabilities
entering into the right side of (6.6). After rescaling (6.8) we find that (6.7)
is given by 
$$\eqalign{\alpha^{4-2D} p^{2-D} q^{2-D} \int^{\alpha N} dn\; n^{D-1} n^{(2D-4)}
n^{2(2-D)}}~~,\eqno{(6.9)}$$
where we show the explicit $n$ dependence of the different factors in (6.7).
Thus, the right side of the inequality (6.6) vanishes for large $\alpha$ as
$\alpha^{4-D}$.

More generally, for the case of $K$ independent random walks we find that the
right side of (6.6) is proportional to
$$\eqalign{\alpha^{(2-D)K} \int^{\alpha N} dn~ n^{D-1}=\alpha^{(2-D)K+D}}~~.
\eqno{(6.10)}$$
Thus, the probability that $K$ random walk paths intersect vanishes if
$$\eqalign{D>{2K\over K-1}}~~.\eqno{(6.11)}$$
This reproduces the well-known condition for a self-interacting scalar quantum
$\phi^{2K}$ field theory to be nontrivial. It gives the critical dimension $D=4$
for $\phi^4$ and $D=3$ for $\phi^6$ field theories.

\bigskip\bigskip\bigskip\bigskip
{\centerline{\bf ACKNOWLEDGMENTS}}
\bigskip
CMB thanks the Institute for Theoretical Physics at the Technion for its
hospitality. MM thanks the BSF and the Gutwirth Fund for partial support. We
thank the U. S. Department of Energy for funding this research.
\vfill\eject

\centerline{\bf REFERENCES}
\bigskip
\item{$^1$} C.~M.~Bender, S.~Boettcher, and L.~Mead, J.~Math.~Phys.~{\bf 35},
XXXX (1994).
\medskip
\item{$^2$} C.~M.~Bender and K.~A.~Milton, J.~Math.~Phys.~{\bf 35}, XXXX (1994).
\medskip
\item{$^3$} The case $R_n=n$ is special because it represents a $D$-dimensional
random walk in {\sl flat} space. We can think of curved space as having the
property that the surface area of the consecutive nested spheres does not grow
like $n^{D-1}$.
\medskip
\item{$^4$} This formula is generic and holds for a random walk on any kind of
lattice. See C.~Itzykson and J.-M.~Drouffe, {\sl Statistical Field Theory}
(Cambridge University Press, Cambridge, 1989), Vol. 1.
\medskip
\item{$^{5}$}
See, for example, C.~M.~Bender and S.~A.~Orszag, {\sl Advanced Mathematical
Methods for Scientists and Engineers} (McGraw-Hill, New York, 1978).
\medskip
\item{$^6$}
By virtue of the initial condition (1.5) and the even-odd parity structure of
the partial difference equation (1.4), the nonvanishing values of $C_{n,t}$ form
a checkerboard pattern; when $t$ is even $C_{n,t}$ vanishes for all even $n$
and when $t$ is odd $C_{n,t}$ vanishes for all odd $n$. Of course this
checkerboard pattern is not apparent in the asymptotic approximations in (2.22),
(2.26), (2.34), and (2.36) because these formulas were derived in the limit as
$t\to\infty$ and therefore they are insensitive to the initial condition (1.5)
given at $t=0$.
\medskip
\item{$^{7}$}
The basis for the reciprocity property may be found in H.~S.~Wall, {\sl Analytic
Theory of Continued Fractions} (Van Nostrand, New York, 1948), p. 281.
\medskip
\item{$^{8}$}
See, for example, R.~J.~Creswick, H.~A.~Farach, and C.~P.~Poole, Jr.,
{\sl Introduction to Renormalization Group Methods in Physics} (John Wiley \&
Sons, New York, 1992), Sec.~1.4.
\medskip
\item{$^{9}$}
See I.~S.~Gradshteyn and I.~M.~Ryzhik, {\sl Table of Integrals, Series, and
Products} (Academic, New York, 1965), p.~949.
\medskip
\item{$^{10}$}
{\sl Ibid.}, pp.~375-377.
\medskip
\item{$^{11}$}
See M.~Aizenman, Phys.~Rev.~Let.~{\bf 47}, 1 (1981) and Commun.~Math.~Phys.~{
\bf 86}, 1 (1982); J.~Fr\"ohlich, Nucl.~Phys.~B {\bf 200} [FS4], 281 (1982) and
J.~Fr\"ohlich, in ``Progress in Gauge Field Theory,'' edited by G.~'t Hooft,
A.~Jaffe, H.~Lehmann, P.~K.~Mitter, I.~M.~Singer and A.~Stora, p.~169 (Plenum,
New York, 1984).
\vfill\eject

\centerline{\bf FIGURE CAPTIONS}
\bigskip

\noindent
 Figure 1. Three-dimensional plot of $C_{n,t}$ on an $(n,t)$ lattice for the
case $D=3$. For a given value of $n\geq 1$, $n-1$ is the smallest value of $t$
for which $C_{n,t}$ is nonzero. (This is because the random walker cannot reach
$region~n$ until time $n-1$.) We find that when $D>1$, $C_{n,t}$ at first
increases with $t$, reaches a maximum, and then dies off to $0$ as $t$ continues
to increase. Also, for fixed $t$ as we vary $n$, we again observe a local
maximum. Thus, in the plot of probabilities there is a crest; this crest lies on
a parabola: $t(D-1) =n^2$. The parabola is indicated on the graph. The graph
shows that the crest is very broad.
\medskip
\noindent
 Figure 2. Same as in Fig.~1 except that $D=1/2$. When $0<D<1$ the maximum value
of $C_{n,t}$ for each fixed value of $t$ lies along the $t$ axis; it no longer
lies on a parabolic curve. As $t\to\infty$ with $n$ fixed, $C_{n,t}$ tends to 
zero very slowly.
\medskip
\noindent
 Figure 3. A plot of $C_{1,t}$ (solid curve) and the leading asymptotic
approximation to $C_{1,t}$ in (2.27) (dashed curve) for $t=2,~4,~6,~8,
\ldots,~3000$ for the case $D=1/2$. The asymptotic approximation clearly becomes
more accurate as $t$ increases.
\medskip
\noindent
 Figure 4. Same as in Fig.~1 except that $D=-1$. When $D<0$ we observe a new
behavior not present in Figs.~1 and 2; namely, that as $t\to\infty$ with $n$
fixed, $C_{n,t}$ approaches $C_{n,\infty}$, a nonzero time-independent function
of $n$. The function $C_{n,\infty}$ is given in (2.36).
\vfill\eject
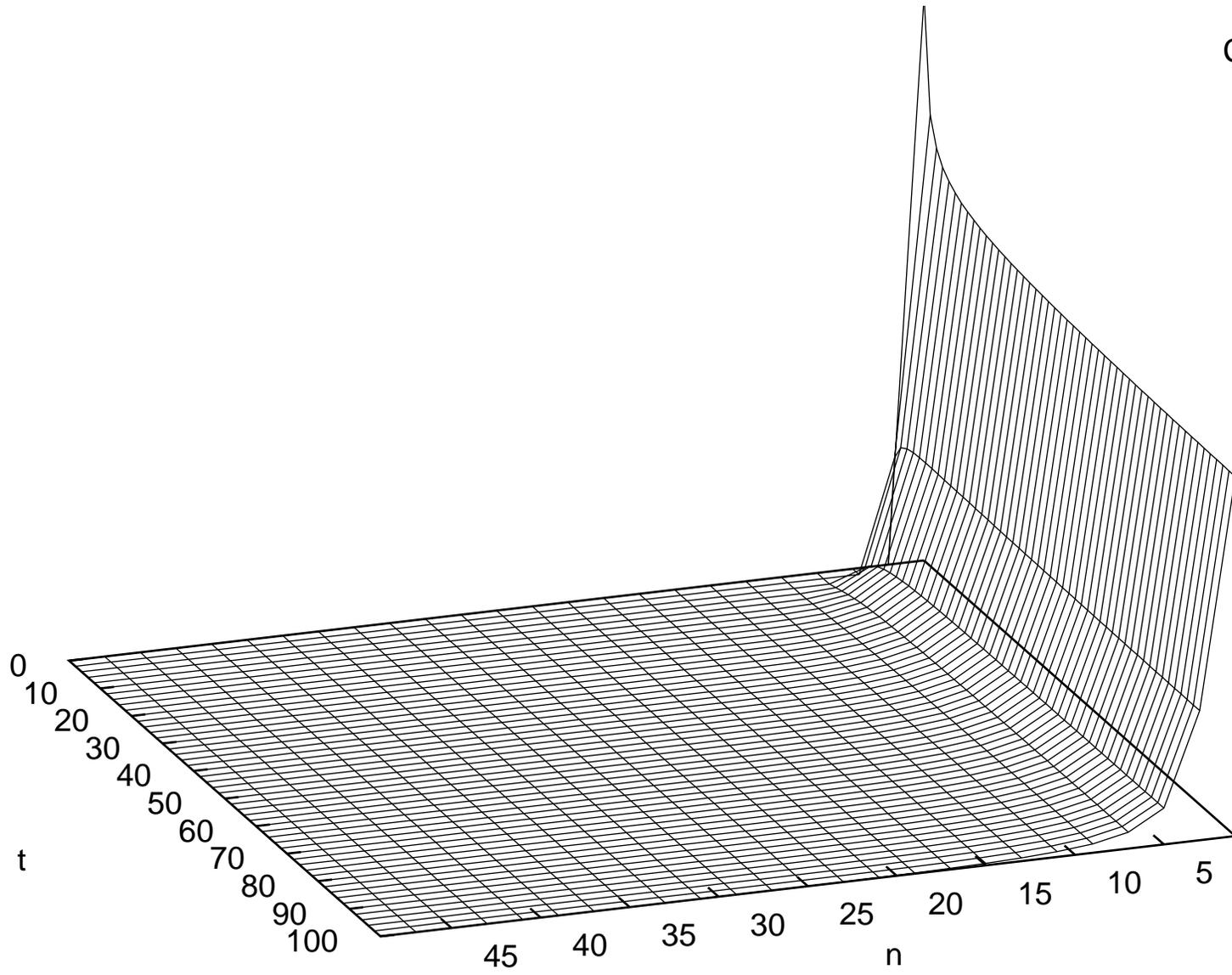
\bye